\documentclass[aps,pra,groupedaddress,twocolumn,showpacs]{revtex4}
\usepackage{graphicx}
\usepackage{pstricks}

\begin{document}
\title{Doppler-free Adiabatic Self-Induced Transparency}
\author{Yu. Loiko$^{a,b}$, C. Serrat$^{a,c}$, R. Vilaseca$^{a}$, V. Ahufinger$^{d,e}$,
J. Mompart$^{d}$, and R. Corbal\'{a}n$^{d}$}

\affiliation{$^a$ Departament de F\'{\i}sica i Enginyeria Nuclear, 
Universitat Polit\`{e}cnica de Catalunya, Colom 11, E-08222 Terrassa, Spain } 
\affiliation{$^b$ Institute of Physics,
National Academy of Sciences of Belarus, Nezalezhnasty Ave. 68, 220072 Minsk, Belarus}
\affiliation{$^c$ Departament de Tecnologies Digitals i de la Informaci\'{o}, Universitat de Vic, E-08500 Vic, Spain}
\affiliation{$^d$ Departament de F\'{\i}sica, 
Universitat Aut\`{o}noma de Barcelona, E-08193 Bellaterra, Spain } 
\affiliation{$^e$ ICREA - Instituci\'{o} Catalana de Recerca i Estudis
Avan\c{c}ats, Barcelona, Spain } 

\begin{abstract}

We demonstrate that a Doppler broadened two-level medium can be made transparent
to a laser pulse by an appropriate adiabatic variation of the laser
field amplitude and its nominal detuning.  
This new technique of adiabatic self-induced transparency (ASIT)
is compared with the well known self-induced transparency (SIT) phenomenon, 
showing that the adiabatic method is much more robust 
to variations of the system parameters. 
We also discuss a possible experimental 
implementation of ASIT using ${^{87}}$Rb atoms.

\end{abstract}

\date{\today }
\maketitle


\section{\label{Introduction}INTRODUCTION}

The ability to tailor the optical properties of a medium 
in such a way that
it becomes transparent to a resonant laser field while its index of
refraction is shaped at wish has been 
used for a wide range of applications,
e.g., from slow light \cite{slowlightcold,slowlighthot} 
and dark-state polariton physics \cite{polaritons} 
to light-matter interfaces \cite{interfaces}. These investigations are
mainly motivated by the fast development of quantum information technologies 
and have been carried out in different types of media from atomic vapours, 
either cold \cite{slowlightcold} or at room temperature \cite{slowlighthot}, 
to solid state materials \cite{storedlightSolid,slowlightSolid}, 
photonic crystals \cite{bookPCs,slowlightPCs} and 
optical fibers \cite{slowlightfiber}. 

For practical implementations it would be desirable 
to work at room temperature but often this implies 
the presence of inhomogeneous broadening that 
tends to reduce the transparency. 
Explicitly, in hot atomic vapors, the atomic motion leads 
to Doppler broadening due to the velocity-dependent shift 
of the laser field frequencies. 
If so, one direct approach to avoid inhomogeneous 
broadening consists of using cold atomic samples although 
it involves demanding technological requirements. 
In fact, albeit there are some proposals on how to reduce 
the negative role of the Doppler broadening in a hot vapor \cite{doppler}, 
a fully satisfactory solution is still challenging. 
Here we will focus on avoiding the adverse effects 
of the Doppler broadening to obtain on-resonance transparency 
for two-level hot atomic vapours. However, our results could 
be extended to other inhomogeneously broadened systems, 
e.g., solids with non-uniform crystalline fields \cite{scully}. 

In a two-level system, there exists a simple approach for achieving
Doppler-free transparency named {\it self-induced transparency} (SIT) \cite{SIT}, 
which consists in shaping a resonant optical pulse propagating in the 
medium such that an integer number of Rabi oscillations is performed. 
This occurs in the coherent regime whenever the pulse area is an integer multiple of $2 \pi$. 
Therefore, SIT requires a very precise temporal control of the Rabi frequency. 
Analogously to SIT, when the pulse area is a multiple of an odd number of $\pi$, 
self-induced absorption (SIA) occurs. 

An alternative approach to achieve resonant transparency consists 
in using induced atomic coherences in three-
or multi-level atomic systems, e.g., by means of the coherent population
trapping (CPT) \cite{cpt_review} and the electromagnetically induced transparency (EIT) 
\cite{eit_review} phenomena or via a double-STIRAP process \cite{stirap_review}. 
In most cases, an intense laser field, the so-called control field, couples 
to an atomic transition while modifying the optical properties of the medium 
for a weak laser field, the probe, acting on an adjacent transition. 
However, the fact that all these atomic coherence effects, CPT, EIT, STIRAP
and double-STIRAP, operate in the so-called two-photon (or Raman) resonance 
condition means that, in the presence of Doppler broadening, 
the two laser fields must have near identical frequencies 
which severely restricts the range of applications 
of atomic coherence effects in hot atomic vapors \cite{doppler}.

In this paper, we present a novel approach that gives rise to Doppler-free 
transparency in a two-level (2L) system resembling a
double-STIRAP process in a three-level (3L) system. 
To achieve self-transparency we will appropriately 
modify in time the amplitude and detuning of the laser pulse. 
It will be shown that this technique, which 
we name {\it adiabatic self-induced transparency} (ASIT), operates 
for a larger parameter range than SIT and is much more robust 
against fluctuations of the parameter values. 

The paper is organized as follows: in Section \ref{Model} the nature of the
ASIT phenomenon is described in the framework of the topological 
equivalence between density matrix equations for a 2L system and the 
Schr\"{o}dinger equation for a 3L system. In Section \ref{ASIT}, the time 
evolution of the 2L system under conditions for ASIT is studied in 
the inhomogeneously (Doppler) broadened case.
Section \ref{SIT-ASIT} is devoted to the comparison
between the proposed ASIT process 
and the already known SIT phenomenon.


\section{\label{Model} Adiabatic Self Induced Transparency}

\begin{figure}[t]
\begin{center}
\includegraphics[scale=0.62,clip=true,angle=0]{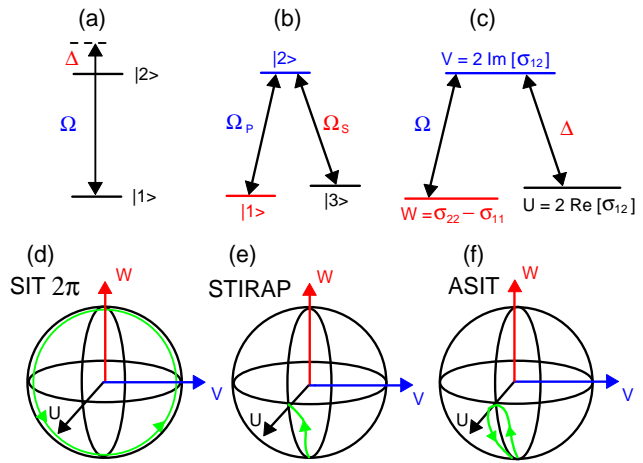}
\end{center}
\caption{ (Color online) 
(a)Two-level system under consideration where $\Omega$ and $\Delta$ are the Rabi frequency 
and the field-atom detuning; (b) Three-level system interacting with two fields with Rabi frequencies 
$\Omega_P$ and $\Omega_S$; (c) equivalence between the density matrix variables of 
the two-level system and the three-level picture. 
Schematic representation, for the two-level case, 
of the Bloch vector evolution 
on the Bloch sphere of the $\left( U, V, W \right)$ variables,
under (d) SIT with $2 \pi$ pulses, (e) STIRAP 
and (f) ASIT processes.}
\label{f1}
\end{figure}

We consider a two-level atom (Fig. \ref{f1} (a)) interacting 
with a classical electromagnetic
field (EM), $\vec{\epsilon}(t)=\vec{e} E(t) \cos (\omega t)$ in the electric dipole approximation. 
Under the rotating wave and slowly varying envelope approximations, 
the coherent dynamics of the two-level atom is given by the
following density matrix or optical Bloch equations:
\begin{equation}
\frac{d}{dt}\left( 
\begin{array}{c}
U \\ 
V \\ 
W
\end{array}
\right) =\left( 
\begin{array}{ccc}
0 & \Delta & 0 \\ 
-\Delta & 0 & -\Omega \\ 
0 & \Omega & 0
\end{array}
\right) \left( 
\begin{array}{c}
U \\ 
V \\ 
W
\end{array}
\right) ~,\quad  \label{Eqs_2L}
\end{equation}
where we have introduced the real-valued variables that define the
coordinates of the Bloch vector moving in a 3D space 
\[
U=2%
\mathop{\rm Re}%
\left( \sigma _{12}\right) ~,\quad V=2%
\mathop{\rm Im}%
\left( \sigma _{12}\right) ~,\quad W=\left( \sigma _{22}-\sigma
_{11}\right) ~,\quad 
\]
$\sigma _{ii}$ and $\sigma_{ij}$  with $i,j=1,2$ account for 
the corresponding populations and coherences.
$\Delta \equiv \omega -\omega _{21} $ is 
the frequency detuning of the field with respect to
the instantaneous atomic transition frequency $\omega _{21}$. 
$\Omega (t) \equiv -\vec{d}_{12} \vec{e}E\left( t\right) /\hbar $ denotes 
the time dependent Rabi frequency with 
$\vec{d}_{12}$ being the electric dipole moment of the atomic transition. 
The Rabi frequency,
without loss of generality, is taken real and positive. 
Throughout the paper, we will consider that
both the detuning and the optical field envelopes are time dependent. 
We name the temporally shaped detuning as {\it D pulse}, 
and the temporally shaped amplitude field as {\it EM pulse}.

Vitanov {\it et al}. \cite{Vitanov2006PRA73} have recently 
shown that Eqs.~(\ref{Eqs_2L}) are topologically
equivalent to the Schr\"{o}dinger equation for a three-level atom 
in a $\Lambda $ configuration interacting 
with two resonant laser fields (Fig.~\ref{f1} (b)) under 
the following identifications: 
$U\Leftrightarrow \overline{C}_{3}$, 
$V\Leftrightarrow \overline{C}_{2}$, 
$W\Leftrightarrow \overline{C}_{1}$, and 
$2\Delta \left( t\right) \Leftrightarrow \Omega_{S}\left( t\right) $, 
$2\Omega \Leftrightarrow \Omega _{P}$ ,
where $\overline{C}_{1,3}=C_{1,3}$ and 
$\overline{C}_{2}=-iC_{2}$, with $C_{i}$
being the probability amplitude to find a three-level atom in state $\left| i \right>$. 
$\Omega _{P,S}$ denote the Rabi frequencies of the two applied fields. One should
remark that for a 2L system the variables $U,V$ and $W$ are real quantities,
while for a 3L system the variables $\overline{C}_{i}$ are, in general, complex numbers.
The analogy between the 2L and the 3L systems is shown in 
Fig.~\ref{f1}(c).

In a 3L system, STIRAP is a widely used technique 
for transferring the atomic/molecular population between 
the two lower levels of a $\Lambda$-scheme and 
it is based on adiabatically following one 
of the three energy eigenstates of the system 
 given by $\left| \Phi (t) \right> = \cos \Theta (t) \left| 1 \right> - %
\sin \Theta (t) \left| 3 \right> $, with the mixing angle defined as
$\Theta ( t ) = \arctan \left[ \Omega_{P} (t) / \Omega_{S} (t) \right]$.
In a 2L system and according to the previous analogy, 
the equivalent to the former energy eigenstate is the following 
combination of the population difference and the real part of the atomic coherence
$ \phi (t) = \cos \theta (t) \times W(t) - \sin \theta (t) \times U(t)$,
with the mixing angle being now 
$\theta ( t ) = \arctan \left[ \Omega ( t ) / \Delta ( t ) \right]$.
Indeed, this analogy opens the possibility to directly extend the results 
that were known previously for 3L systems, obtained from 
the adiabatic temporal variation of the Rabi frequencies of the two coupled fields,
to 2L systems, by appropriately engineering the temporal profile of the detuning and 
the Rabi frequency of a single EM pulse.
We note that the temporal variation of the field amplitude and the detuning can be
accomplished for instance by chirped pulses, and 
that chirped pulse excitation to achieve population inversion 
in a 2L system has been discussed via 
Stark Chirped Raman Adiabatic Passage (SCRAP) \cite{2001VitaonARP_review}.

Such analogy between STIRAP and chirped two-state excitation has been studied
in \cite{Vitanov2006PRA73} in the complete coherent case, e.g., 
without spontaneous emission, and in the absence of Doppler broadening.
From that study it follows that for the 2L-STIRAP case the Bloch vector
moves from the south pole ($W=-1$, $U=0$ and $V=0$) to the equator 
($W=0$, $U=1$ and $V=0$) [see Fig.~\ref{f1}(e)] on the Bloch sphere. 
Starting with the population in the lower level and by
applying first a D pulse and later on, and with an appropriate overlap, the EM pulse,
the system ends up in a coherent superposition of the two states. 
In this process, the variable $V$ accounts for EM field absorption 
(positive values) and stimulated emission (negative values),
 $U$ describes the dispersion properties 
of the transition, and $W$ gives the transfer of population. 

In a 2L system, a functional definition of transparency is that 
the two-level atom evolves coherently from $W(t_{initial})=-1$ (population 
in the lower level) back to $W(t_{final})=-1$. 
This coherent evolution should be fast enough 
such that spontaneous emission becomes negligible, 
which in turn, means that there is not irreversible EM absorption.
Such definition of transparency is equivalent to state that 
the time averaged imaginary part of the atomic coherence 
weighted by the Rabi frequency is equal to zero, 
i.e., $g\left\langle  \Omega V   \right\rangle = 0$, where $g=\frac{N {\omega}_{12} {d^2_{12}}}{2 \hbar {\epsilon}_0 c}$ is the gain parameter per unit length with $N$ the atomic density, ${\epsilon}_0$ the electric permitivity and $c$ the speed of light. Hereafter the brackets $\left\langle ... \right\rangle$ mean time averaging.
 
As it has been mentioned above, a well known method to obtain transparency in a 2L system is the 
SIT phenomenon \cite{SIT}, i.e., coherent induced absorption of pulse energy during 
the first half of the pulse followed by coherent induced emission of the same 
amount of energy back into the beam direction during the second half of the pulse. 
In this case, the complete transfer of the population from the lower to the upper state 
and back can be realized with a fully resonant excitation 
whenever the pulse area is a multiple of $2\pi$.
Pulse areas multiple of an odd number of $\pi$ result in complete absorption (SIA). 
The SIT/SIA phenomena have
an absorptive nature, which means that in the Bloch sphere the evolution occurs predominantly 
in the plane $\left(V,W\right) $ with $U=0$ [see Fig.~\ref{f1}(d)], 
i.e. the transfer of the population is due to the EM absorption and emission 
processes involved and, 
therefore, $g\left\langle  \Omega V   \right\rangle = 0$ with  
$g^2\left\langle  ( \Omega V )^2   \right\rangle %
/ \left\langle  \Omega^2 \right\rangle ^2 \neq 0$.

The novel method to obtain Doppler-free 
transparency in a 2L system, 
which we propose here, i.e. ASIT, consists in a process 
that resembles a double-STIRAP sequence in a 3L system.
In a 3L system (Fig.~\ref{f1}(b)), a double-STIRAP process starts and ends 
in the ground state $\left| 1 \right>$ since 
the mixing angle starts at $\Theta = 0$, goes to $\Theta = \pi /2$, and 
ends back again at $\Theta = 0$.
To produce a double-STIRAP process, a first Stokes pulse has 
to be applied at the beginning of the process, 
then a partially overlapping pump pulse, 
and finally a partially overlapping second Stokes pulse. 
Alternatively, a relatively longer Stokes pulse 
that overcomes the pump pulse can be also used.  
In a similar manner, ASIT can be produced in a 2L system by applying 
an EM pulse in between two D pulses.
The ASIT phenomenon is mainly of dispersive nature,
since the evolution occurs predominantly 
in the plane $\left( U,W\right) $, with 
much weaker relative excitation of the variable $V$ related with absorption 
and of variable $W$
[see Fig. \ref{f1} (f)] and, therefore, 
$g\left\langle  \Omega V   \right\rangle = 0$ and 
$g^2\left\langle  ( \Omega V )^2   \right\rangle %
/ \left\langle  \Omega^2 \right\rangle ^2 $ takes values smaller than in the SIT case.

Up to now we have not considered the incoherent processes 
that might be present in the system. 
In fact, there are no analogies between the 2L and  3L systems
regarding incoherent processes like spontaneous emission or dephasing. 
In the 3L case, the spontaneous decay rate of the excited state does 
not play any role either in the STIRAP or in the double-STIRAP processes, 
because the whole adiabatic process 
occurs without coupling the excited state. 
In contrast, in the 2L case, spontaneous emission plays an important role because 
the exited state is always involved in the process. 
We will assume that the population difference decays 
at a rate $\gamma$ while the decay of the real 
and imaginary parts of the coherence is $\Gamma/2$. 
In general, $\Gamma$ accounts for several incoherent processes 
like spontaneous emission, collisions, dephasing or phase fluctuations.  
Here we only consider the effect of the spontaneous emission 
in the decay of the coherences and therefore $\Gamma=2 \gamma$. 
The requirement of full coherence for ASIT to work imposes a maximum 
time duration of the process 
limited by the lifetime of the levels, i.e., $T \gamma \ll 1$, 
where $T$ is the total time for the ASIT process. 
\begin{figure}[t]
\begin{center}
\includegraphics[scale=0.7,clip=true,angle=0]{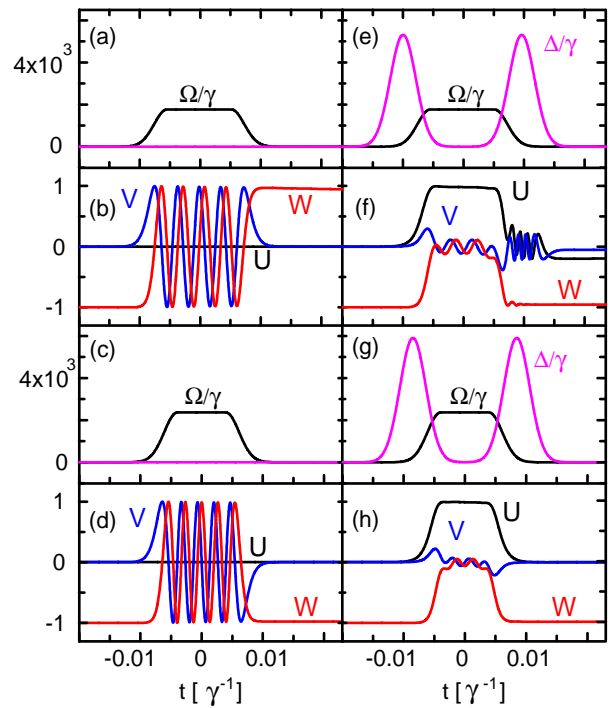}
\end{center}
\caption{
(Color online)
Temporal profiles of the EM (black line) and D (magenta line) pulses (a,c,e,g).
Time evolution of the population difference ($W$) and the 
medium coherence $\left( U,V \right)$ (b,d,f,h) in a 
homogeneously broadened two-level system 
under the SIA (a,b), SIT (c,d) and ASIT (e-h) processes.
The areas of the pulses are equal 
to $A=A_\Omega=A_\Delta=9\pi$ (a,b,e,f) and 
$=10\pi$ (c,d,g,h). 
The D pulses and the rise and fall parts of the EM pulses
have gaussian profile.
The temporal width of the D pulses
is $\protect\tau_{\Delta} = 0.003 \gamma^{-1} $ , $\Delta_{0} = 0$. 
$\overline{\Delta } / \gamma =5318$ (e,f) and $=5909$ (g,h).
$\Omega_{0} / \gamma =1772$ (a,b,e,f) and $=2363$ (c,d,g,h).
The rise $\tau_{r}$ and fall $\tau_{f}$ time of the EM pulses is equal to
$\tau_{r,f} = \protect\tau_{\Delta} /2 $.
The D pulses are shifted away
by $\Delta \tau = 1.5 \tau_{\Delta}$ from the top flat part of the EM pulse.
The time duration of the top flat part of the EM pulses is set
to keep the EM pulse area indicated above.
In all cases represented, the detuning of the EM pulses 
during the top flat part is close to zero value.
}
\label{f2new}
\end{figure}
Under these conditions, comparison between
the three just above described processes of
SIA, SIT and ASIT is shown in Fig.~\ref{f2new}
where the time evolution of the variables $\left( W,U,V \right)$
in a homogeneously broadened 2L medium is represented.
It should be mentioned that in the case we consider here 
Figs.~\ref{f2new} (b,d,f,h) represent
the temporal response of a single 2L atom as well as
of a homogeneously broadened medium composed of such 2L atoms.
In contrary to the SIA and SIT processes (see Figs.~\ref{f2new} (a-d))
where the final state of the system strongly depends on the EM pulse area, 
under the ASIT phenomenon 
the population returns back to the lower level,
giving rise to the transparency of the 2L medium,
independently on the area of the exciting EM pulse
as demonstrated in Figs.~\ref{f2new} (e-h).
Although the strength of the light-atom interaction is larger for the SIA and SIT processes
than for the ASIT process, but it is strong in all cases.


\section{\label{ASIT}ASIT in inhomogeneously broadened media}

Now we focus on  the role of the Doppler broadening 
in the 2L system under consideration by assuming a Maxwell distribution for the projection
of the atomic velocities on the light propagation direction.
We will solve the ASIT process for different atomic velocity classes 
and sum up the total response along the Maxwell distribution. 
For an atom moving with some velocity $v$ along the direction 
of the propagating laser field (with wavenumber $\mbox{k} = \omega / c $), 
the instantaneous field detuning 
will be $\Delta \left( v ,t \right) = \Delta \left( t \right) - kv$ , 
where $\Delta \left( t \right)$ is the nominal detuning for atoms at rest. 
For definiteness we choose temporal Gaussian profiles 
for the D and the EM pulses:
\begin{eqnarray}
\Delta \left( t\right) &=&\Delta _{0}+\overline{\Delta }\left(
e^{ -\left( t-\Delta \tau \right)^{2} /\tau _{\Delta } ^{2} } +
e^{ -\left( t+\Delta \tau \right)^{2} /\tau _{\Delta } ^{2} } \right)
~,\quad  \nonumber \\
\Omega \left( t\right) &=&\Omega _{0} e^{ -t^{2}/\tau _{\Omega }^{2} }
~,\quad \label{Eqs_Pulses}
\end{eqnarray}
where $\overline{\Delta }$ defines the amplitude of the
temporal variation of the detuning with respect 
to the constant detuning offset $\Delta _{0}$, 
$\Omega _{0}$ is the amplitude of the EM pulse,
$\tau _{\Delta , \Omega } $ are the time durations of the D and EM pulses,
and $\pm \Delta \tau $ are the delays of the D pulses 
with respect to the center of the EM pulse. 

For a complete transfer of the atoms, the double-STIRAP process should be 
performed adiabatically, condition that  
for atoms moving with velocity $v$ reads \cite{Vitanov2006PRA73}:
\begin{eqnarray}
T \sqrt{\Omega_0^2+ \Delta^2 \left( v ,t \right)} > 10 
~.\quad \label{Eqs_AdiabaCond}
\end{eqnarray}

Let us assume, for simplicity and without losing generality,
that $\Delta_0$ and $\overline{\Delta}$ are both positive. 
Then for the homogeneous case or for a small Doppler width,
i.e. for $\Delta \omega_{D} < \Delta$, the adiabaticity condition 
(\ref{Eqs_AdiabaCond}) can be fulfilled for all atoms simultaneously, 
provided that appropriate values for the parameters, 
in particular for $\Omega_0$, are chosen.
This might not be the case, however, for hot atomic vapours, 
where  the velocity-dependent Doppler shift can strongly modify the D pulse offset 
during the ASIT process for some atomic velocity classes. 
The largest reduction in the instantaneous values of the 
detuning $\Delta^2 \left( v ,t \right)$ will occur 
for atomic velocities verifying $kv>\Delta_0$ and 
in particular for $kv \sim (\Delta_0+\overline{\Delta })$, where 
the detuning will remain close to zero during significant time intervals 
(those where the D pulse passes near its maxima values). 
This can break the adiabaticity condition (\ref{Eqs_AdiabaCond}) for atoms 
with such velocity, so that 
they can be transferred to the excited state
breaking the transparency under the double-STIRAP process.
One needs to note that to maintain adiabaticity of the process for small detuning
$\Delta \left( v,t \right)$ the Rabi frequency $\Omega_{0}$ 
should be increased substantially, see for instance \cite{Vitanov2006PRA73}.

\begin{figure}[t]
\begin{center}
\includegraphics[scale=0.9,clip=true,angle=0]{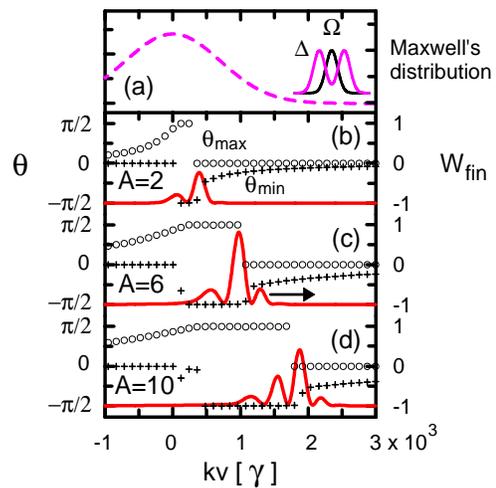}
\end{center}
\caption{ (Color online) 
Distributions for the final value (red solid line) 
of the
atomic
population inversion, $W_{fin}$,
and of the maximum (empty circles) 
and minimum (crosses) 
values for the mixing angle $\theta$ 
with respect to the Doppler shift $kv$ at different 
amplitudes of the EM and D pulses (b-d). 
The areas of the EM and D pulses, defined as
$A_\Omega = \Omega_{0} \tau_\Omega/\sqrt{2}$ and
$A_\Delta = \overline{\Delta} \tau_\Delta/\sqrt{2}$, are equal to each other 
$A = A_{\Omega} / \pi = A_{\Delta} / \pi $ = 2 (b), = 6 (c), = 10 (d).
The Maxwell distribution of the atomic velocities is shown in (a) (dashed line). 
The inset in (a) presents the time evolution of the exciting EM ($\Omega$) and
D ($\Delta$) pulses. Abrupt decrease in the $\theta_{max}$ corresponds to the 
velocity class for which $kv \sim \Delta_0 + \overline{\Delta } $.
The temporal widths of the input pulses 
are $\protect\tau_{\Delta} = \protect\tau_{\Omega} = 0.01 \gamma^{-1} $ , 
$\Delta_{0} = 0$ and $\Delta \tau = 1.5 \tau_{\Omega}$. 
The Doppler width is fixed at $\Delta \omega_{D} = 10^{3} \gamma$ and $\Gamma = 2 \gamma$.
Macroscopic response of the inhomogeneously broadened 2L medium
is given by the integration of the atomic response
[shown by red solid lines in figures (b-d)]
weighted by the probability density for the atomic velocities
of the Maxwell distribution 
[shown by the dashed line in figure (a)].
}
\label{f7v3}
\end{figure}

\begin{figure}[t]
\begin{center}
\includegraphics[scale=0.75,clip=true,angle=0]{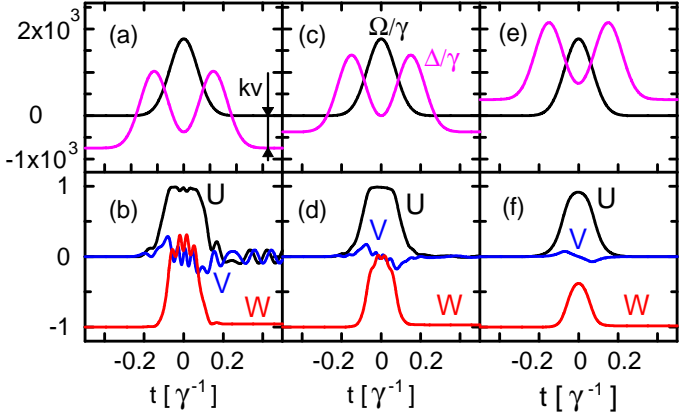}
\end{center}
\caption{
(Color online) 
Temporal profiles for the EM pulse (black line) and
the total detuning seen by the atom (magenta line),
for three different atomic velocity classes (a,c,e):
$kv = \Delta \omega_{D} 3 /4$ (a,b),
$kv = \Delta \omega_{D} /3$ (c,d) and
$kv = - \Delta \omega_{D} /3$ (e,f).
Figs. (b), (d) and (f) represent the time evolution of
the atomic variables $W$, $U$ and $V$.
The pulse areas are $A_\Omega=A_\Delta=10\pi$. 
The other parameters are the same as in Fig.~\ref{f7v3}
}
\label{f3new}
\end{figure}

We have studied in detail the response of 2L atoms 
with different values of the Doppler shift $kv$ 
along the Maxwell distribution
by exciting them with D and EM pulses 
of the form given in (\ref{Eqs_Pulses}). 
Some of the results are presented in Figs.~\ref{f7v3} and \ref{f3new},
for a specific Doppler frequency profile and 
a specific temporal pulse profile [see Fig.~\ref{f7v3} (a)].
Fig.~\ref{f3new} shows the total atom-field detuning 
(sum of the D-pulse value plus the Doppler shift $kv$) [see, Figs.~\ref{f3new} (a), (c) and (e)],
and the atomic-state evolution for 
three different atomic velocity classes $v$ [see, Figs.~\ref{f3new} (b), (d) and (f)].
It is clear from Fig.~\ref{f3new} that in all cases the light-atom interaction is quite strong
[it is less strong in the case of Figs.~\ref{f3new} (e) and (f) 
because in that case the total field detuning is largest], 
but always the final state of the population difference $W$ is the same 
as the initial state, i.e., $W=-1$.
This is due to the adiabaticity of the ASIT process.
The dependence of the final value of 
the
atomic
population inversion, $W_{fin}$, 
i.e. the response of a single 2L atom,
with respect to the Doppler shift $kv$
is presented in Figs.~\ref{f7v3} (b-d), red solid line, 
for different values of the area of D and EM pulses 
at a constant temporal duration of the pulses.
It is clearly shown in the figure that, 
in the velocity domain pointed out above, 
with $kv \sim (\Delta_0+\overline{\Delta})$ , the 2L atoms are strongly perturbed 
by the pulses and they do not end at the initial state $W=-1$.


\begin{figure}[t]
\begin{center}
\includegraphics[scale=0.9,clip=true,angle=0]{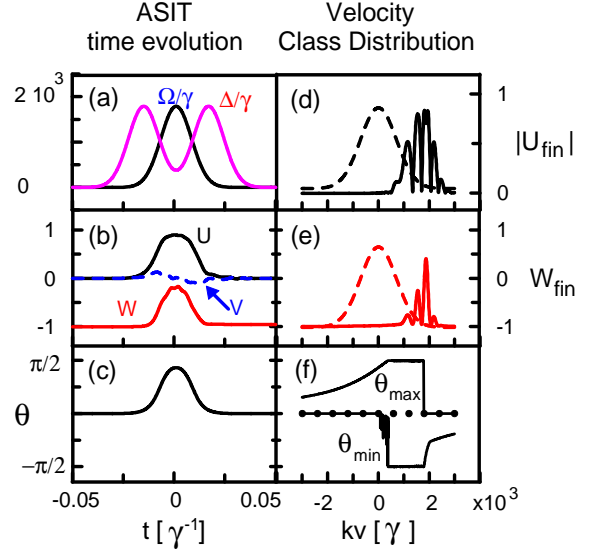}
\end{center}
\caption{ (Color online) Time evolution of the
macroscopic
population difference ($W$) and the 
medium coherence $\left(U,V) \right)$ (b) in the 
inhomogeneously broadened two-level system 
under excitation of EM and D pulses shown in (a). (c) shows the evolution 
of the mixing angle, $\theta$, for atoms 
with zero projection of the velocity onto the light propagation direction. 
The pulse areas are $A_\Omega=A_\Delta=10\pi$ .
Distribution of the final amplitude for the
single atom
variables
$U$ and $W$ on the Doppler shift $kv$ is 
shown in (d) and (e), respectively. Dashed lines in (d) and (e) are plotted 
to visualize the Maxwell distribution of the atomic velocities. The maxima (upper line) 
and minima (lower line) for the variable $\theta$ are shown in (f), 
where dots represent its zero final value.
The other parameters are the same as in Fig.~\ref{f7v3}}
\label{f3}
\end{figure}

Therefore, the macroscopic response of the inhomogeneously broadened 2L medium considered
is provided predominantly by atoms in the velocity domain of nonadiabatic response.
As known the macroscopic absorption of atoms within a particular velocity domain of the Doppler profile
is defined by absorption of individual atoms
[represented by $W_{fin}$, see red solid lines in Figs.~\ref{f7v3} (b-d) and Fig.~\ref{f3} (e) ]
weighted by the probability density for these atoms to belong to this particular velocity domain
within the Maxwell distribution (shown by the dashed lines in Fig.~\ref{f7v3} (a) and Fig.~\ref{f3} (e) ).
Therefore, the absorption of the 2L medium considered could be reduced either
by decreasing the absorption of individual 2L atoms or
by moving the velocity domain of nonadiabatic response to the wings of the Maxwell distribution
where the probability density for atomic velocities is strongly reduced.
We found that by increasing the amplitude $\overline{\Delta}$ of the D pulse 
(increasing simultaneously the amplitude of the EM pulse 
to keep the same pulse areas) the domain of nonadiabatic response 
could be shifted to the wing of the Doppler profile,
as it is clearly shown in Figs.~\ref{f7v3}(b-d).
In this way its negative role is reduced 
due to the much lower relative atomic population 
at the wing of the Maxwell distribution shown by the dashed line in Fig.~\ref{f7v3}(a), 
almost all the atoms undergo an adiabatic process 
and, consequently, the medium becomes essentially transparent.

In order to verify the previous statement, we present 
in Fig.~\ref{f3} further results corresponding to the case considered 
in Fig.~\ref{f7v3}(d), which describes 
an inhomogeneously broadened system of 2L atoms excited by D and EM pulses 
of the form given in (\ref{Eqs_Pulses}) with area of $A=10 \pi$. 
Fig.~\ref{f3}(a) shows the sequence of 
the exciting D and EM pulses in such case. 
The time evolution of the
macroscopic
variables $\left( U, V, W \right)$
(obtained at each time step by the summation of single atom response
weighted in accordance with the Maxwell distribution of the atom velocity)
and of the mixing angle $\theta$ (the latter is presented only for atoms 
with $kv = 0$) for the ASIT process is plotted in Fig.~\ref{f3}(b) 
and Fig.~\ref{f3}(c), respectively. One 
can see that the ASIT phenomenon has predominantly a dispersive nature, 
since the quantity $V$ takes values 
much smaller than those taken by $U$ [Fig.~\ref{f3}(b)].
In Fig.~\ref{f3} the amplitude of the D pulses 
has been chosen two times larger 
than the Doppler width of the inhomogeneous transition. 
In such a case, the spectral components of the EM pulse induce almost 
identical evolutions in 2L atoms from different velocity 
classes of the Maxwell distribution
(except for the perturbed domain mentioned above 
at $kv \sim (\Delta_0+\overline{\Delta })$ ). 
Indeed, we found that a given exciting pulse in the inhomogeneously broadened 2L system 
will produce almost the same response that would be induced 
in a homogeneously broadened 2L system.
It should be mentioned that for small values of the amplitude of the D pulse,
the perturbed domain is moved to the center of the Maxwell distribution,
as shown in Fig. \ref{f7v3}.
This strongly modifies the response of the inhomogeneously broadened 2L atoms
because large part of them are transferred to the excited state, 
making the ASIT process ineffective.


\section{\label{SIT-ASIT} ASIT versus SIT}

\begin{figure}[t]
\begin{center}
\includegraphics[scale=0.95,clip=true,angle=0]{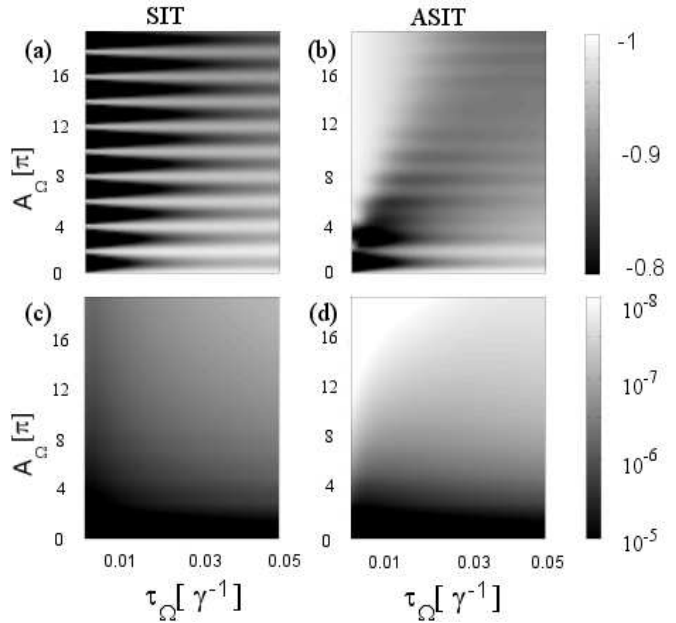}
\end{center}
\caption{
Contour plot of the final values of 
the
macroscopic
population inversion, $W_{fin}$, after SIT (a) and ASIT (b) processes
in a Doppler broadened 2L system 
as a function of the pulse area and 
duration. $A_{\Delta} = A_{\Omega}$ in (b) and (d), 
and $A_{\Delta} = 0$ in (a) and (c).
Other parameters of the 2L system 
and of the pulses are the same as in Fig.~\ref{f7v3}.
White regions in (a) and (b) correspond to 
the domains where $W_{fin} \rightarrow -1$.
Plots for time averaged values of 
$g^2\left\langle (\Omega V )^2 \right\rangle / %
\left\langle  \Omega^2 \right\rangle ^{2}$ in $g^2$ units
indicate the
pure
absorptive nature of the SIT phenomenon (c) and the
more dominantly dispersive
nature of the ASIT process in the domain
close to the left vertical axis (d).}
\label{f5}
\end{figure}

In this section we compare the robustness of the ASIT technique
with respect to the SIT one by studying the effect of the variation 
of the pulse parameters.
We consider the evolution of an inhomogeneously (Doppler) broadened 2L system. 
Fig. \ref{f5} shows the numerical results for the final
values of the
macroscopic
population difference $W_{fin}$ [Figs. \ref{f5} (a) and \ref{f5} (b)]
and for the time averaged quantity 
$g^2 \left\langle ( \Omega V )^2 \right\rangle / \left\langle  \Omega^2 \right\rangle ^2 $ describing 
the absorptive character of the process [Fig.~\ref{f5}(c) and Fig.~\ref{f5}(d)], 
in the parameter plane defined by the 
area of the D and/or EM pulses and the pulse width. 
Fig.~\ref{f5}(a) and Fig.~\ref{f5}(c) correspond to a SIT process 
and Fig.~\ref{f5}(b) and Fig.~\ref{f5}(d) correspond to an ASIT process.
 
Since the ASIT process is based on a double-STIRAP technique, 
and since the STIRAP mechanism is more effective 
when the exciting pulses have equal amplitudes, 
we have assumed that D and EM pulses have equal areas, 
$A_\Omega = A_\Delta$, and time durations
in the simulations of Fig. \ref{f5}.
Moreover, it has been assumed that the delay time 
between consecutive D and EM pulses 
is $\Delta \tau=1.5 \tau$, for which it is known that 
STIRAP transfer by Gaussian pulses is most efficient \cite{stirap_review}.
For the SIT process we have considered only 
an EM pulse of area $A_\Omega$ ($A_\Delta=0$).

White regions in Fig.~\ref{f5}(a) and Fig.~\ref{f5}(b) correspond to 
the domains with $W_{fin} \rightarrow -1$, which 
means that the final state is the same as the initial one 
and thus there is transparency. 
Fig.~\ref{f5}(a) shows that to achieve transparency in the SIT case 
the pulse area needs to be precisely controled (multiples of $2\pi$), 
especially for short pulses. 
But even for these particular values of the pulse area, 
Doppler broadening
and relaxation processes
prevent from reaching complete transparency 
(the best values in Fig.~\ref{f5}(a) are obtained for pulse area 
equal to $2\pi$ and short pulse duration, 
where $W_{fin} \sim -0.973$ at $\tau_{\Omega} = 2 \times 10^{-3} \gamma^{-1}$). 
For longer pulses (and thus with smaller peak amplitude) 
the broadening due to Doppler effect (and also due to spontaneous emission) 
is more present because its relative influence on the value of 
the generalized Rabi frequency increases. 
In contrast, Fig.~\ref{f5}(b) shows that in the ASIT case transparency is feasible 
with pulses of arbitrary area, provided this area is larger than a value 
(approximately $4\pi$) which is almost equal to that
required by the adiabaticity condition (\ref{Eqs_AdiabaCond})
in the double-STIRAP process. Note also in Fig.~\ref{f5}(b) 
that the degree of transparency is larger for 
smaller pulse durations (i.e., near the vertical axis), 
where values very close to $-1$ ($W_{fin} \sim -0.995$ at $\tau_{\Omega} = 2 \times 10^{-3} \gamma^{-1}$
for pulses of area $A_{\Omega} = A_{\Delta} = 10 \pi $
) are achieved. For larger pulse durations (and thus lower pulse amplitudes) 
it is progressively more difficult to push the velocity classes not fulfilling 
the adiabaticity condition (i.e., classes $kv \sim (\Delta_0 + \overline{\Delta } )$) 
to the wings of the Maxwell distribution and larger pulse areas are required for that. 
Also the influence of spontaneous emission is larger for longer pulses. 
Note also in Fig.~\ref{f5}(b) that near the horizontal axis there is a second region 
with values of $W_{fin}$ close to -1. Such region does not correspond to pure ASIT, 
since the adiabaticity condition is not fulfilled; 
rather, it can be seen as a reminiscence of the SIT effect 
due to the EM pulse (compare with Fig.~\ref{f5}(a)), 
combined with the relaxation  processes. Fig.~\ref{f5}(c) and Fig.~\ref{f5}(d), confirm the previous discussion. 
The absorptive character is smaller in the ASIT case 
than in the SIT case.

\begin{figure}[t]
\begin{center}
\includegraphics[scale=0.95,clip=true,angle=0]{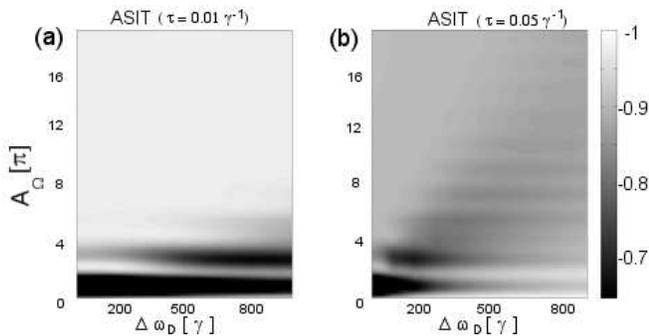}
\end{center}
\caption{ Contour plot of 
macroscopic variable
$W_{fin}$ under conditions of ASIT in the plane 
$\left( A_{\Omega} = A_{\Delta} , \Delta \omega_{D} \right)$ for 
a Doppler broadened 2L system of $^{87}Rb$ atoms 
excited by EM and D pulses of equal areas
for two values of the time duration 
$\tau_{\Omega} = \tau_{\Delta} = 0.01 \gamma^{-1}$ (a)
and $\tau_{\Omega} = \tau_{\Delta} = 0.05 \gamma^{-1}$ (b).
The considered transition of the $^{87}Rb$ atoms is 
$(\left|{5 ^{2}S_{1/2},F=2,m_F=2}\right\rangle $
$\leftrightarrow$
$\left|{5 ^{2}P_{3/2},F=3,m_F=3}\right\rangle $
with $\lambda_{0} = 780$ nm, $\gamma = 5.83 \times 2 \pi \times 10^6$ $s^{-1}$ 
and $\Gamma = 2 \gamma$.
}
\label{f6v4}
\end{figure}

It is known that the Doppler width in optical transitions can be modified by
cooling or heating the atoms. In such cases it is possible to change 
the Doppler width in a large range of values from below 
to few orders above the homogeneous width.
In order to show the feasibility of ASIT for real hot atoms 
we have investigated the effects of the Doppler width in specific atomic transitions. 
The results are summarized in Fig.~\ref{f6v4}, where we plot the 
dependence of $W_{fin}$ in a hot gas of $^{87} Rb$ atoms 
as a function of the amplitudes of the EM and D pulses 
and of the Doppler width of 
the inhomogeneously broadened transition 
$(\left|{5 ^{2}S_{1/2},F=2,m_F=2}\right\rangle $
$\leftrightarrow$
$\left|{5 ^{2}P_{3/2},F=3,m_F=3}\right\rangle $,
for two values of the pulse width $\tau_{\Omega} = \tau_{\Delta} = 0.01 \gamma^{-1}$ (Fig.~\ref{f6v4}(a)) 
and $= 0.05 \gamma^{-1}$ (Fig. \ref{f6v4}(b)). 
It is clear from Fig. \ref{f6v4} that 
nonadiabatic effects are negligible for pulses with
large area and short time duration. 
In contrast they are pronounced for long pulses 
and also for short pulses with small area, when the 
amplitude $\overline{\Delta}$ 
of the D pulse becomes comparable to the Doppler width of the transition.
By taking into account that at room temperature ($T=300 ^{0} {\rm K}$),
the Doppler width is equal to $\Delta \omega_{D} = 530 \gamma$ 
we can conclude, [see Fig. \ref{f6v4}(a)], 
that the reported phenomenon of ASIT 
can be feasible experimentally in an atomic vapour cell of rubidium
at room temperature and under excitation of nanosecond EM and D pulses 
with pulse area of few $\pi$.

In conclusion, we have demonstrated the feasibility of the adiabatic self 
induced transparency (ASIT) technique, which yields
Doppler-free transparency for electromagnetic pulses in a 2L system. 
Our technique consists of driving the 2L system with a 
laser pulse whose detuning with respect to the atomic transition  
and whose EM field amplitude are temporally varied in a proper shape. 
The detuning modulation can be introduced either 
by using the Stark or the Zeeman effect to obtain 
time-dependent energy levels in the atom/molecule, 
which gives rise to time-dependent detunings, or 
alternatively by introducing a suitable chirp on the EM pulse. 
We have studied the robustness of the reported phenomena against 
modifications of the pulse parameters and
we have shown that the ASIT process is much more robust than the
conventional SIT, in the sense that the area of the pulses
needs not to be adjusted with precision 
and it can be much less sensitive to Doppler broadening. 
Because of the adiabaticity condition, 
the pulse areas for ASIT need in general 
to be somewhat larger than for SIT, 
for which a $2\pi$ area is sufficient, 
but it could be implemented with 
no essential difficulty in real Doppler-broadened atomic systems 
such as for instance rubidium hot vapors.
Although the results presented in this article have been obtained for 
Gaussian pulses we have also checked 
the validity of the ASIT technique 
for pulses with other forms of temporal profiles.

\begin{acknowledgments}
We acknowledge support from 
the Spanish Ministry of Education and Science (MEC)
under the \textit{Programa Ram\'on y Cajal} 
and under the contracts FIS2005-07931-C03-03, FIS2005-01497 and FIS2008-02425, 
from the project "Quantum Optical Information Technologies" 
within the Consolider-Ingenio 2010 program 
under the contract CSD2006-00019. 
Support from the Catalan Government 
under \textit{CRED}-program and 
under the contracts SGR2005-00358 and SGR2005-00457
is also acknowledged.
\end{acknowledgments}

\end{document}